\documentclass[sigconf]{aamas}  
\usepackage{booktabs}
\usepackage{times}  
\usepackage{helvet}  
\usepackage{courier}  
\usepackage{url}  
\usepackage{graphicx}  
\usepackage{amsmath}
\usepackage[ruled,vlined,linesnumbered]{algorithm2e}
\usepackage{multirow}
\usepackage{booktabs}
\usepackage{mathrsfs}
\usepackage{amssymb}
\usepackage{mathrsfs}

\begin{document}

\title{The Price of Governance: A Middle Ground Solution to Coordination in Organizational Control}  




\author{Chao Yu}
\affiliation{%
  \institution{School of Computer Science and Technology, Dalian University of Technology}
  \streetaddress{Dalian University of Technology}
  \city{Dalian, 116024}
  \state{China}
  \postcode{116024}
}
\email{cy496@dlut.edu.cn}

\begin{abstract}  
Achieving coordination is crucial in organizational control.  This paper investigates a middle ground solution between decentralized interactions and centralized administrations for coordinating agents beyond inefficient behavior. We first propose the {\it price of governance} (PoG) to evaluate how such a middle ground solution performs in terms of effectiveness and cost. We then propose a hierarchical supervision framework to explicitly model the PoG, and define step by step how to realize the core principle of the framework and compute the optimal PoG for a control problem. Two illustrative case studies are carried out to exemplify the applications of the proposed framework and its methodology. Results show that by properly formulating and implementing each step, the hierarchical supervision framework is capable of promoting coordination among agents while bounding administrative cost to a minimum in different kinds of organizational control problems.

\end{abstract}

%


\maketitle


\section{Introduction}
Imagine a fairy tale that you were the king of an ancient kingdom. Owning a large piece of land, you were now facing a tricky problem: how to divide the land into smaller administrative districts such that you can run your kingdom as efficiently as possible? You might choose to directly govern each citizen by yourself. While you could take full control of your kingdom  in this way, a heavy burden would definitely be imposed. Or, you might prefer relieving yourself by letting each district administrate itself. However, as the number (size) of districts is getting larger (smaller), your kingdom tends to be more fragmented and thus your authority is prone to be weakened. Facing this dilemma, you were puzzled: what is the best size for the partition such that your control is not impaired but at the same time the whole kingdom can function efficiently?

Although this fairy tale seems naive, it reveals a fundamental yet challenging issue in organizational control, where \emph{centralization} and \emph{decentralization} are two completely opposite solutions to guarantee system performance. Application domains include but are not limited to the management of supply chains \cite{giannoccaro2018centralized}, resource allocation in cognitive radio networks \cite{hasegawa2014optimization} or multiuser OFDMA networks \cite{yassin2017centralized}, and multi-robot formation/consensus control \cite{oh2015survey,yu2014collective}. In these domains, it is crucial to design efficient \emph{coordination mechanisms} that enable all the agents to reach an agreement in areas of common interest. While \emph{centralized} mechanisms often rely on a dictatorial authority to formulate, specify and enforce how the agents should coordinate with each other, \emph{decentralized} mechanisms enable agents to coordinate via local interactions and without relying on any centralized authority.

On one hand, it is generally believed that system performance could be improved upon given dictatorial control over agents' actions. Imposing such control, however, can be costly or even infeasible in large systems due to the expense of high administrative cost. Moreover, as the environments where agents reside in become even more dynamic and open, continuously monitoring and governing each agent's behavior will soon become intractable. On the other hand, pure decentralized mechanisms usually cannot guarantee satisfactory performance if no external interventions or explicit mechanisms are imposed. As in the \emph{social dilemma} scenarios, for example, pure rational behavior based on best-response reinforcement learning will end up with non-cooperative defection, also known as \emph{selfish equilibria}, which is suboptimal with respect to the social welfare \cite{bazzan2011learning}. In \emph{coordination games}, high level of coordination among distributed learning agents are rarely achieved if no extra mechanisms are introduced, especially in situations with stochastic and partial information \cite{kapetanakis2002reinforcement}. Similar results can also be observed in various kinds of \emph{congestion games}, such as resource allocation problems, in which inductive reasoning or regret-based learning algorithms would always converge to an inefficient equilibrium that is far worse than the optimum \cite{oh2008few}.

In this paper, we investigate the possibility of a middle ground solution between decentralized interactions and centralized administrations, by which self-interested agents can utilize a proper level of coordination to improve their performance beyond inefficient equilibria or uncoordinated behavior. Based on the two quantitative criteria of \emph{price of anarchy} (PoA) and \emph{price of monarchy} (PoM), we propose the {\it price of governance} (PoG) to evaluate how such a middle ground solution performs in terms of effectiveness and cost. The PoA measures the \emph{inefficiency} of a decentralized solution with respect to the natural objective function, and can be defined as the ratio of the objective function value of a decentralized solution at its convergence to that of an optimum achieved by a fully centralized solution. Corresponding to the PoA, PoM is defined as the practical \emph{cost} of centralized administration. By combining these two criteria into an overall value of PoG using a combination function, an optimal middle ground solution can be properly discovered to make the best trade-off between \emph{inefficiency}/\emph{cheapness} of decentralization and \emph{optimality}/\emph{high cost} of centralization.

Motivated to explicitly model PoG and seek its optimal value, we then introduce a hierarchical supervision framework, which nicely reflects the features of organizational structure and hierarchical governance in human societies. We define step by step how to realize the core principle of hierarchical supervision in the framework and compute the optimal PoG for a control problem. We then carry out a preliminary set of simulations in two case studies: the \emph{norm learning} (NL) problem \cite{delgado2002emergence} and \emph{multi-agent resource selection} (MARS) problem ~\cite{oh2008few}, to evaluate the efficacy of our framework. The two chosen cases are typical coordination problems that have been widely and extensively studied before. The existence of \emph{sub-norm} in the NL problem and the \emph{rationality paradox} in the MARS problem are both famous puzzles that have challenged researchers for decades. Results show that the hierarchical supervision framework can facilitate coordination among agents (i.e., reducing the PoA) compared to a pure decentralized solution. At the same time, an optimal PoG can be achieved to bring out the maximum coordination promotion while bounding the PoM significantly lower than that of a centrally administrated system.

The remainder of the paper is organized as follows. Section~\ref{sec:methodology} introduces PoG and the proposed supervision learning framework. Section~\ref{sec:case_studies} presents two case studies. Section~\ref{sec:related_work} discusses related work. Finally, Section~\ref{sec:conclusions} concludes this chapter.

\section{Seeking the Optimal PoG}\label{sec:methodology}
In this section, we formally define the concept of PoG, which is a measure of a middle ground solution between centralized control and decentralized interactions. We then introduce the hierarchical supervision framework and formulate step-by-step implementation of seeking the optimal PoG for a coordination problem.
\subsection{The Price of Governance}\label{sec:PoG}

In order to measure the degradation of social welfare due to selfish behavior of players, the \emph{price of anarchy} (PoA) was introduced in \cite{koutsoupias1999worst} for makespan minimization in scheduling games. In its original form, PoA was defined as the worst-case ratio between the value of social cost in a \emph{Nash equilibrium} and that of some \emph{social optimum}. Along this definition, researchers have done extensive investigation on the theoretical bound analysis of PoA under various conditions of games \cite{wang2016analysis,roughgarden2015intrinsic}, and applied the PoA concept in a variety of real-world domains, including routing in transportation, resource allocation of network bandwidths, communication network design, competitive dynamic pricing, and supply chain management~\cite{chen2012design}.

Game theoretical analysis of PoA has deepened our understanding of the impact of selfish behavior on system performance. To expand the realm of its applications, PoA has also been generalized to many other contexts, such as for defining the inefficiency of a \emph{multiagent learning} algorithm in resource allocation problems~\cite{oh2008few}, and more broadly the inefficiency of decentralization \cite{youn2008price,cole2015decentralized}. Following this, we define PoA as the ratio of social cost at an equilibrium (convergence) to the optimal social cost that could possibly be achieved by a centralized optimization approach. More formally, \emph{Price of Anarchy} (PoA) can be given by:
\begin{equation}\label{equ:poa}
PoA = \frac{\psi_{opt} - \psi_{dis}}{\psi_{opt}}, \quad PoA \in [0,1]
\end{equation}
where $\psi_{opt}$ is the optimal performance using a centralized solution and $\psi_{dis}$ is the performance of a decentralized solution. The performance can be any criterion that evaluates a solution, e.g., coordination level or convergence speed. A lower PoA indicates a more effective decentralized solution.

Analogous to the price of anarchy, we can define the price of monarchy as the practical cost of maintaining centralization in a system. To simplify illustration, we mainly discuss managerial cost in terms of communication cost. Thus, the lower bound of the price of monarchy is found in a fully decentralized non-communicating system, and the upper bound of the price of monarchy is found in a fully centralized system. Let $\varphi_{dis}$ and $\varphi_{opt}$ denote a communication cost function of a decentralized solution and a centralized solution, respectively. \emph{Price of Monarchy} (PoM) is given by:
\begin{equation}\label{equ:pom}
PoM = \frac{\varphi_{dis}}{\varphi_{opt}}, \quad PoM \in [0,1]
\end{equation}

For a target problem and a given coordination solution, PoA and PoM can be calculated in a closed form or estimated using specific modelling and simulation techniques. The \emph{Price of Governance} (PoG) then can be computed using a combination function $\Gamma$ of PoA and PoM:
\begin{equation}\label{equ:pog}
PoG =  \Gamma(PoA,PoM)
\end{equation}

By defining different function $\Gamma$, one can capture various patterns of behavior towards totalitarianism or liberalism,  depending on specific purpose of solving a target problem.

\subsection{The Hierarchical Supervision Framework}
Inspired to model PoG and seek its optimal value, we then introduce the hierarchical supervision framework, which is composed of the following five steps.

\textbf{Step 1: \emph{Segmentation of social groups}.}
To model hierarchical supervision and organizational governance in human societies, a social group is first divided into a set of subgroups according to some predefined methods. Interactions of sub-group members are purely local and decentralized, and may be constrained by certain external factors such as network topologies or social relationships. In each sub-group, a superior governor monitors and administrates the behavior of its subordinates. The governor can be any one of the subordinate agents in the sub-group or another dedicated agent. A governor can also interact with another governor or other governors to exchange their information or learn from each other using social learning strategies.

\textbf{Step 2: \emph{Aggregation of public opinions}.}
In each sub-group, agents make decisions in a fully decentralized manner. Agents may
\begin{itemize}
  \item learn from the outcome of interaction with another randomly chosen member using reinforcement learning;
  \item copy another member's behavior using some imitation rules;
  \item simply make decisions independently.
\end{itemize}
Each agent then reports its decision to its governor, who then aggregates all the information from its subordinates to form a public opinion using democratic mechanisms. This public opinion summarizes the overall attitude towards the members' behavior in the governor's group.

\textbf{Step 3: \emph{Generation of supervision policies}.}
After obtaining the public opinion, each governor then generates a supervision policy by exchanging its information with another governor and learning from situations in other subgroups. The generated supervision policy is deemed as the most successful behavior in the whole society, and can be in different forms such as being the majority action adopted by the members, the highest rewarded action or their combinations.

\textbf{Step 4: \emph{Adaption of local behavior}.}
The supervision policy is then passed down to the group members by the governor in order to entrench its influence in the group. According to the targeted problem, this integration process can be conducted in distinct manners. The supervision policy can be used to dictate the policy for group members directly, or as a suggestive guide to adapt members' behavior through modifying their behaviorial parameters (e.g., learning speed or exploration mode), transforming the environmental components (e.g., states or rewards), or changing the way how members interact with each other (e.g., to whom to interact).



\textbf{Step 5: \emph{Calculation of the optimal PoG}.} PoA can be computed as the ratio of the performance value at convergence to the optimal performance value using a centralized solution that a single governor supervises the whole group. PoM can be computed as the practical communication cost of maintaining centralization in the group. The communication cost can be in different forms such as the number of message exchange or the geometrical distance between group members and the governor. Then, PoG can be calculated using a predefined combination function $\Gamma$.

As the number or size of subgroups indicates different levels of centralization and thus different PoG\footnote{More subgroups means smaller size of each subgroup, and thus the whole group is more decentralized. In the extreme case when subgroup size is 1, each agent governs its own behavior and this indicates a fully decentralized case.}, the whole problem is then reduced to seeking the optimal size of subgroups in which case the minimal PoG can be achieved. Let $p\in \mathcal{P}$ be a partition of the group. The problem is now transformed into the following optimization problem:

 \begin{align}\label{equ:optimization}
& \min_{p\in \mathcal{P}}\;  PoG=\Gamma(PoA,PoM)\nonumber\\
&s.t.  \enspace  \mathcal{\daleth}_p(PoA,PoM)=0,  \quad  \forall p\in \mathcal{P}
\end{align}
where $\mathcal{\daleth}$ is the PoA and PoM relationship function that is determined by the target problem and the chosen coordination solution. The constraints in Eq.~(\ref{equ:optimization}) means that for each partition of the group, its PoA and PoM value should satisfy their relationship function. For a problem that PoA, PoM and their relationship function can be computed in a closed form, general optimization methods can be applied to compute the solution of this optimization problem. The other more straightforward way is to sample in the partition space in different levels of granularity, and then apply simple approximation methods to estimate the relationship between PoA and PoM. The optimal PoG  then can be easily derived by solving the combinatorial equations of relationship function $\daleth$ and PoG function $\Gamma$.

\section{Two Case Studies}\label{sec:case_studies}
This section provides two case studies to illustrate the applications of the proposed framework and its methodology. We show that by properly formulating and implementing each step, the hierarchical supervision framework is capable of promoting coordination while bounding administrative cost to a minimum in different kinds of organizational control problems.
\subsection{The Norm Learning Problem}\label{sec:norm_learning}
\emph{Social norm} is an important concept in multiagent systems to facilitate coordination among agents by posing constraints on agents' behavior~\cite{shoham1997emergence}. The \emph{norm learning} (NL) problem deals with how a social norm can be established in a bottom-up manner via agents' local learning interactions. This problem has attracted a great interest in recent years and extensive investigations have been conducted under various assumptions about agent interaction protocols, societal topologies, and observation capabilities \cite{hao2017efficient,vouros2017learning,hasan2015fast,airiau2014emergence,tang2018optimal}.
\subsubsection{Problem Description}
Considering a typical setting of network topology, a group of agents are organized in a social network and each agent can only interact with its neighbors, using either reinforcement learning approaches or some predefined imitation rules. The interactions between two agents can be modeled as a pure \emph{Coordination Game} (CG) \cite{sen2007emergence}, in which the agents are rewarded positively when their actions are consistent and penalized otherwise. The goal is to enable all the agents to reach an agreement (\emph{social norm}) in the whole system. Although this problem seems simple, successfully solving it is a challenging task due to the widely recognized existence of \emph{sub-norms}, which prevents the full convergence of a consistent social norm in the whole group \cite{mihaylov2014decentralized}.
\subsubsection{Application of the Methodology}
We now provide an illustration of how to apply the proposed methodology in solving NL problem in a structured system where agents interact with each other using basic reinforcement learning approach, particularly Q-learning algorithm.

\textbf{\emph{Step 1: Group segmentation.}} We use an $R*R$ grid network by default, and separate it into $n*n$ $(1\leq n \leq R)$ subgroups\footnote{In case of $n$ being not divisible by $R$, the remaining agents on the border are included in a single subgroup.}, each of which is denoted as $C_x$. We imagine a governor located in the geometrical center of each subgroup.

\textbf{\emph{Step 2: Aggregation of public opinion.}} At time step $t$, in each subgroup $C_x$, agent $i$ chooses an action $a_i$ with the highest Q-value or randomly chooses an action with an exploration probability $\epsilon_i^t$. Agent $i$ then interacts with a random neighbor $j$ and receives a payoff $r_i$. The learning experience in terms of action-reward pair $(a_i, r_i)$ is reported to agent $i$'s governor $x$, and the governor aggregates all the information from its subordinates into two values $F_x$ and $R_x$. Value $F_x(a)$ indicates the overall acceptance (i.e.,~frequency) of action $a$ in subgroup $C_x$ and value $R_x$(a) indicates the overall reward of action $a$ in $C_x$. $F_x(a)$ can be calculated as $F_x(a) = \sum\limits_{i \in C_x} \delta(a,a_i)$, where $\delta(a,a_i)$ is the Kronecker delta function, which equals to 1 if $a=a_i$, and 0 otherwise. $R_x(a)$ can be calculated by $R_x(a) = \frac{1}{F_x(a)}\sum\limits_{i \in C_x, a_i = a} r_i$. Especially, $R_x(a)$ is set to 0 if $F_x(a) = 0$. Each governor $x$ then combines the actions of each subgroup into a public opinion $o_x$ using democratic voting mechanism ($o_x = arg~max_a ~F_x(a)$), which means that $o_x$ is simply the action most accepted by the subgroup. 

\textbf{\emph{Step 3: Generation of supervision policies.}}
After generating the public opinion, each governor then generates a supervision policy $a_x$, which indicates the \emph{social norm}, i.e.,~the most successful behavior, in the neighborhood. To this end, the governor resorts to \emph{social learning} with another governor by changing their information and comparing the performance of their public opinions. Many \emph{social learning} strategies can be applied for this purpose, and we here employ the widely used imitation rules from the \emph{evolutionary game theory} (EGT)~\cite{szabo2007evolutionary}, given by $p_{x\rightarrow y} = \frac{1}{{1 + {e^{ - \beta ({u_y} - {u_x})}}}}$, where $p_{x\rightarrow y}$ is a probability for governor $x$ to imitate the action of neighboring governor $y$, ${u_x} = R_x({o_x})$ is the fitness of the public opinion of governor $x$, ${u_y} = R_y({o_y})$ is the fitness of neighboring governor $y$, and $\beta>0$ is a parameter to control selection bias.

\textbf{ \emph{Step 4: Adaption of local behavior.}}
Each agent $i$ in a subgroup adjusts its learning behavior in order to comply with the generated supervision policy from its governor. By comparing its action $a_i^t$ with the supervision policy $a_x$, agent $i$ can evaluate whether it is performing well or not so that its learning behavior can be dynamically adapted to fit the supervision policy. Learning rate and exploration rate are two fundamental tuning parameters in RL. Heuristic adaption of these two parameters thus models the adaptive learning behavior of agents. More specifically, when agent $i$ has chosen the same action with the supervision policy (i.e.,~$a_i^t = a_x$), it decreases its \emph{learning rate} to maintain its current state ($\alpha_i^{t+1}  = (1 - \lambda)\alpha _i^t$), otherwise, it increases its learning rate to learn faster from its interaction experience ($\alpha_i^{t+1}  = (1- {\alpha _i^t})\lambda + \lambda$), where $\lambda \in [0,1]$ is a parameter to control the adaption rate. The \emph{exploration rate} can be updated likewise. Finally, agent $i$ updates its Q-value using the new learning rate $\alpha_i^{t+1}$ and/or exploration rate $\epsilon_i^{t+1}$. The proposed mechanisms are based on the concept of ``winning'' and ``losing'' in the well-known multi-agent learning algorithm WoLF (Win-or-Learn-Fast)~\cite{bowling2002multiagent}. While the original meaning of ``winning'' or ``losing'' in WoLF and its variants is to indicate whether an agent is doing better or worse than its Nash-Equilibrium policy, this heuristic is gracefully borrowed here to evaluate the agent's performance against the supervision policy.


\textbf{\emph{Step 5: Calculation of optimal PoG.}}  We sample the size of subgroup from $n=1$ to $n=R$ and derive PoA and PoM for each size of subgroup through simulations. The PoA indicates the ratio of performance loss at an equilibrium (convergence) to the optimal performance that could possibly be achieved by a centralized optimization approach. In the NL problem, the performance loss can be reflected by the consensus level of the whole group, i.e., the proportion of agents in the whole system that have not achieved a consensus. Thus, PoA can be computed as the proportion of agents with sub-norms in the system. As for the PoM, the geometric distance between a group agent and its governor is used to represent the communication cost\footnote{In terms of number of messages exchanged, decentralized and centralized solutions may have almost the same number of messages to be exchanged. However, centralized solutions can bring about higher level of jam since agents are utilizing the same common channel \cite{oh2008few}}. For each case of subgroup size, the PoA and PoM pair can be obtained accordingly. Function approximation methods then can be applied to fit all the pairs to derive the relationship function $\daleth$ between PoA and PoM. Finally, given a predefined PoG function $\Gamma$, the optimal PoG value and its associated subgroup size can be found accordingly.


\begin{figure}[h]
\centering
\includegraphics[width=0.51\textwidth]{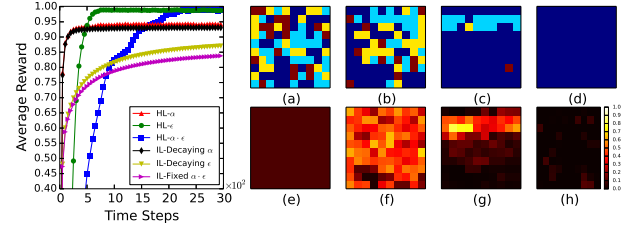}
\centering
\caption{\textbf{The left part} plots the dynamics of PoA in terms of average reward using different learning approaches. HL-$\alpha$, HL-$\epsilon$, and HL-$\alpha\cdot\epsilon$ denote, respectively, the three approaches under the proposed hierarchical supervision framework when agents adapt their learning rate $\alpha$, exploration rate $\epsilon$, and both rates at the same time. The other three approaches represent the IL approaches with a decaying $\alpha$, a decaying $\epsilon$ or a fixed $\alpha$ and $\epsilon$. \textbf{The right part} shows the evolution dynamics of agents' actions and learning rate. Subfigures (a)-(d) give the snapshots of action distribution in the group and (e)-(h) present the snapshot of values of learning rate $\alpha$ at step $t=0, 10, 50, 1000$, respectively.}
\label{fig:overall_performance}
\end{figure}

\subsubsection{Experiments and Results}\label{subsubsec:norm_results}


First, we would like to test whether the proposed hierarchical supervision framework is capable of facilitating coordination among agents (i.e., reducing the PoA), compared on a pure decentralized learning approach. We conduct the investigation on a $10*10$ grid network, which is separated into several $4*4$ clusters (the remaining 2 agents on the border are included in a single group). We consider stateless version of Q-learning, and each agent can choose from 4 actions as default. Parameters $\alpha$ and $\epsilon$ are initially set to 0.1 and 0.01, respectively. Moreover, parameter $\beta$ and $\gamma$ are both set to 0.1. The final results are averaged over 1000 independent runs. We compare our hierarchical learning approaches (denoted as HL) to the individual learning (IL) approach, which is a fully decentralized approach in that agents learn randomly with another agent in the population and update their strategies independently.

The left part of Figure~\ref{fig:overall_performance} shows that the coordination ratio of the whole group using different approaches increases as learning proceeds, but the hierarchical learning approaches (especially HL-$\epsilon$ and HL-$\alpha\cdot\epsilon$) can reach a higher convergence level than the individual learning approaches. This result indicates that by introducing a certain level of centralized control, the PoA can be greatly reduced. As shown from subfigures (a)-(d), the four actions are equally distributed in the population at first (a), but as time proceeds, the action represented by blue color emerges as the dominating social norm in the whole population (d).

\begin{figure}[h]
\centering
\includegraphics[width=0.53\textwidth,height=3.5cm]{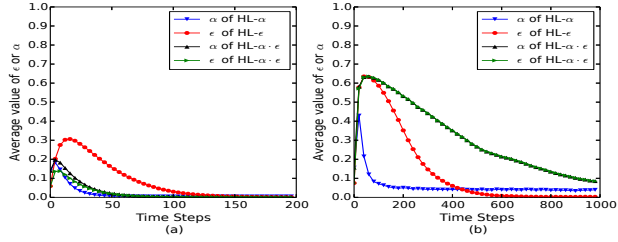}
\centering
\caption{Dynamics of $\epsilon$ and $\alpha$ in case of (a) 2 actions and (b) 4 actions. Dynamics of $\alpha$ and $\epsilon$ with HL-$\alpha\cdot\epsilon$ overlap with each other because of the identical update method.}
\label{fig:4dynamic}
\end{figure}

In order to reveal the behavior dynamics under the proposed framework, subfigures (e)-(h) in Figure~\ref{fig:overall_performance} present the snapshot of values of learning rate $\alpha$ during learning. The values are set to be 0.1 initially (e) and then increase to higher values (f). As time proceeds, the values decrease gradually (g) and finally reach zero (h). At the beginning, failed interactions among agents drive agents to learn fast to escape from this adverse situation. As learning moves on, agents' actions are more and more consistent with the supervision policy, and thus $\alpha$ decreases accordingly. Figure~\ref{fig:4dynamic} further plots the learning dynamics with different action sizes. In a 4-action scenario, the values change more drastically at first and then take a longer time to decrease to zero. This is because the probability to find the right action as the norm is greatly reduced when the number of actions gets larger, and thus the coordination process is greatly hindered.

 \begin{figure}[h]
  \begin{center}
    \includegraphics[width=0.35\textwidth]{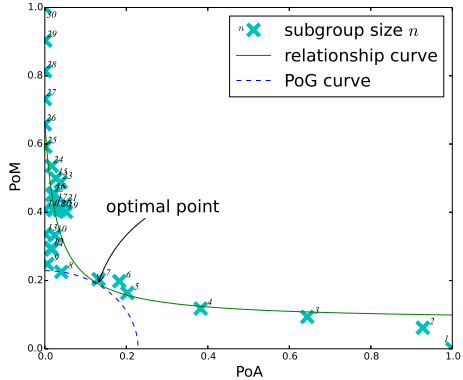}
  \end{center}
  \vspace{-15pt}
  \caption{The relationship of PoA and PoM, and the calculation of the optimal PoG in the NL problem. The relationship function between PoA and PoM is fitted using curve $f(x)=\frac{a}{(x+b)}+c$, with the fitted coefficient a, b, c being 0.0174, 0.0299, 0.0821, respectively.}
  \label{fig:optimal_curve_norm}
  \end{figure}

Figure~\ref{fig:optimal_curve_norm} plots the relationship between PoM and PoA when the subgroup size takes different values of $n$ in a $30\times30$ grid network. As we can see, a larger size can result in a higher consensus level (i.e., lower PoA). This is easy to understand because each governor can have a more powerful control force over the group comparatively when the subgroup size is larger. The communication cost, however, also increases as the subgroup size becomes larger, causing a higher PoM. It is obvious that the PoA and PoM are two contradictory criteria that evaluate the coordination performance. PoA indicates the consensus level while PoM indicates the cost for achieving this performance. Higher PoM indicates a more centralized system (i.e., higher cost) and thus a better coordination performance (i.e., lower PoA) can be achieved. We can observe that the PoA and PoM exhibit a monotonous relationship with a long tail phenomenon. This indicates that the PoA can be reduced significantly by only introducing a bit of centralized control, e.g., when subgroup size is from 2 to 8. This improvement, however, is only at the expense of very low communication cost, as reflected by the low value of PoM. As an illustration to compute the lowest PoG, we simplify the PoG function as $PoG=\sqrt{PoA^2+PoM^2}$. The optimal PoG is then on the tangent point between the curve of PoG function and the relationship function. In the case of $30\times30$ grid network as shown in Figure~\ref{fig:optimal_curve_norm}, the optimal PoG can be achieved when the grid is divided into several $7\times7$ clusters, which means that the whole group will achieve the maximum coordination promotion at the expense of lowest administration cost.

\subsection{The MARS Problem}\label{subsec:bar_learning}
The \emph{multi-agent resource selection} problem (MARS) is a class of \emph{congestion games} characterized by a large number of self-interested agents competing for common resources \cite{oh2008few}. This so called congestion effect is apparent in many real-world situations, ranging from traffic routing in transportation systems, bandwidth allocation in communication networks, to other versions of \emph{tragedy of the commons} that are characterized by negative externalities.
\subsubsection{Problem Description}
Formally, an MARS problem can be defined as a quadruple of $(N, \Theta, A, R)$, in which $N={1,2,...,n}$ is a set of agents, $\Theta={r_1,...,r_m}$ denotes a set of resources available for agents in $N$, $A_t=a_1\times...\times a_n$ denotes the resource choices of the agents at time $t$ where $a_i\in\Theta, \forall i\in N$, and $R_t: \Theta\times A_t\rightarrow \Re$ is a reward function. In MARS, a reward associated with using a resource is defined as a function of the number of concurrent users of the resource, and all users using the same resource share the same reward. So, agents' valuations of congested resources are not exogenously-determined, but rather are endogenous functions of one each other's actions.

The \emph{El Farol bar problem} (EFBP) is a simple example of MARS, which was introduced in \cite{arthur1994inductive}. In EFBP, a set of $n$ agents repeatedly make decisions of whether to attend a bar or not on certain nights. The only observations available to the agents are the past history of attendance at the bar.
The bar is a congested resource so that the payoff of attending a bar is high only if the number of attendees at the bar on the night is less than a certain threshold $\zeta$. The agents receive the worst payoff if the bar is over crowded. Thus, an agent needs to reason about the attendance so as to decide whether to attend it or not. However, a rational agent always fails to learn the best decision based on its expected reward, since all agents are simultaneously learning the same information and reason in the same manner. This leads to the \emph{rationality paradox} in general MARS problems.


\subsubsection{Application of the Methodology}
In the bar problem, we apply the methodology of our hierarchical supervision framework in the following five steps.

\textbf{\emph{Step 1: Group segmentation.}} We imagine 900 people living in a district with 30*30 blocks, which can be separated into several subgroups with $n*n$ blocks. The different values of $n$ thus indicate different levels of centralized governance.

\textbf{\emph{Step 2: Aggregation of public opinion.}} Each member $i$ has a probability of $p_i$ to attend the bar. A community governor in each subgroup collects the information of how many subordinates went to the bar last night (i.e., attendance radio in the subgroup) and the average reward of subordinates in the subgroup\footnote{Actually, more sophisticated methods can be applied here by reasoning on a longer past history of attendance.}. The attendance radio in the subgroup is the public opinion and the average reward is its performance

\textbf{\emph{Step 3: Generation of supervision policies.}} Based on the public opinion from the subgroup in terms of attendance radio and its reward, the governor applies reinforcement learning to update her knowledge about whether to attend the bar or not. To apply the tabular form of Q-learning, we transform the attendance radio between $[0,1]$ into a set of discrete actions. Moreover, the governors need to learn from other governors by comparing their performance. Simple imitation rule (e.g., the Fermi rule) can be employed for this purpose. If the governor accepts the action of another governor in the social learning process, she informs the new action to her subordinates as the subversion policy, otherwise, the governor chooses an action based on the Q values using $\varepsilon$ exploration strategy (i.e., with a high provability of $1-\varepsilon$ to choose the action with the highest Q value and a small provability of $\varepsilon$ to explore randomly), and informs her subordinates the chosen action accordingly.

\textbf{ \emph{Step 4: Adaption of local behavior.}} When receiving the supervision policy in terms of attendance probability $a_i$, a subordinate agent $i$ updates its policy of attendance probability $p_i$ directly by $p_i=(1-\mu)*p_i + \mu*a_i$, where $\mu$ is a learning rate. In the next round, agent $i$ decides whether to attend the bar based on the updated attendance probability.

\textbf{\emph{Step 5: Calculation of optimal PoG.}} We normalize the PoA by $PoA_n=1-\frac{r_n-r_{min}}{r_{max}-r_{min}}$, where $r_n$ is the average reward of group size $n$, $r_{min}$ and $r_{max}$ are the minimum and maximum average reward for different sizes of subgroups, respectively. The calculation of PoM is: $PoM_n=\frac{c_n-c_{min}}{c_{max}-c_{min}}$, where $c_n$ is the communication cost of group size $n$, $c_{min}$ and $c_{max}$ are the minimum and maximum communication cost for different sizes of subgroups, respectively. The geometric distance is still used to represent the communication cost. The relationship function between PoA and PoM, as well as the optimal size and PoG value then can  be derived in the same way as in the NL problem.

\begin{figure}[t]
\centering\includegraphics[width=3.5in]{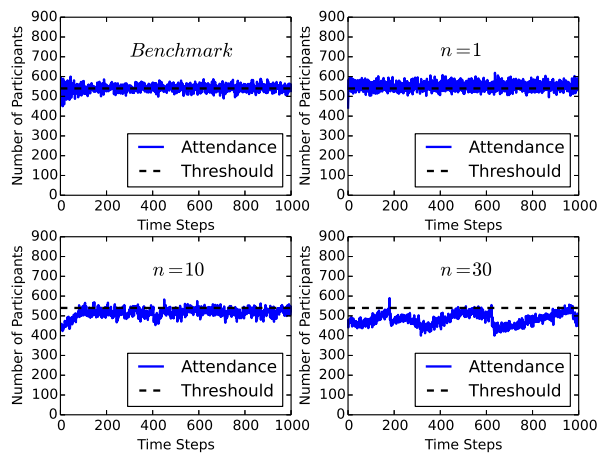}
\caption{The dynamics of attendance in the bar problem. In the benchmark case, agents make their decisions without any hierarchial control. In the hierarchical supervision case, $n$ indicates the size of subgroups in the 30*30 group.}\label{fig:1}
\end{figure}

\subsection{Experiment Results}
We set the threshold of attendance $\zeta$ to 540, which means that the bar can only accommodate $60\%$ of total 900 agents at most. If more than 540 agents attend the bar, they will receive a penalized reward to indicate a crowded situation. The learning parameter $\mu$ is set to 0.1. In Q-learning, the range of attendance probability in between $[0,1]$ is divided into 50 discrete actions, with the interval between two adjacent action being probability of 0.02. Exploration rate $\varepsilon$ is set to 0.01. In each trial, the experiment runs for 1000 nights (i.e., time steps), and we take the average of rewards in the last 100 time steps for evaluation. As a benchmark, agents maintain the probability of attendance, and update the policy by learning with an arbitrary agent in the system using the imitation rule. The results are averaged over 1000 trials.

From Figure \ref{fig:1}, it is clear that the number of participants in the benchmark model is fluctuating along the threshold, indicating an inefficient equilibrium among agents. When an agent predicts the attendance at the bar is lower than $\zeta$, then the agent decides to attend the bar. Since the other agents also reason in the same manner, the entire population decides to attend the bar, ending up with the worst payoffs. Therefore, agents face contradicting outcomes by making decisions based on their rationality. This \emph{rationality paradox} can be greatly alleviated using the proposed framework by imposing a certain level of centralized control on the agents. As can be seen, the number of overcrowded nights is significantly reduced when $n=10$. The performance is further promoted when the governor has a wider control range when $n=30$ in a fully centralized manner.

\begin{figure}[t]
\centering\includegraphics[width=3.5in]{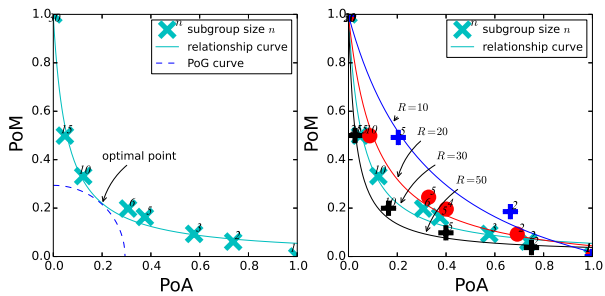}
\caption{The relationship of PoA and PoM in the bar problem in the case of 30*30 group (the left part), and the relationship of PoA and PoM in different sizes of agent groups (the right part)}\label{fig:2}
\end{figure}

The left part in Figure \ref{fig:2} presents the relationship curve when the 30*30 group is divided to different sizes of subgroups. It shows the same pattern of result as the NL problem, and the optimal value of PoG can be computed in the same way as described before. The right part in Figure \ref{fig:2} presents the impact of different sizes of agent population. It is apparent that a higher population size $R$ generates a lower optimal PoG. This is an interesting phenomenon that is a bit of counterintuitive. It demonstrates that the proposed methodology is more suitable for larger systems, where decentralized control methods cannot perform well because of the narrow vision of agents and lack of centralized control. This also provides an explanation on why in real-life situations, large organizations and systems such as countries and companies usually embody the feature of hierarchical supervision structures to make an elegant balance between centralized governance and decentralized administration.

\section{Related Work} \label{sec:related_work}

PoA has been extensively studied in the area of computational economics and computer science to study the inefficiency of selfish behavior \cite{koutsoupias1999worst,andelman2009strong}. The mainstream research in this direction focuses on the computational analysis of upper or lower bound on PoA  under various conditions of congestion games and real-world applications. Wang et al. analyzed PoA in the sequential repeated congestion games \cite{wang2016analysis}. Feldman et al. developed an analytical framework for bounding PoA in fundamental classes of games when there is a large number of players \cite{feldman2016price}. The PoA has also been analyzed in decentralized transportation networks \cite{youn2008price} and price mechanisms for network resource allocation \cite{chen2012design}. However, most of these studies do not consider an explicit cost associated with the decision making process, which is unrealistic in real-life problems where manageable cost or communication cost is inevitable in sustaining the global system order. While the work in \cite{oh2008few} has extended the definition of PoA as a measure of inefficiency of a multiagent learning algorithm in MARS problems, and considered adminstration cost in such contexts, our work highlights a general hierarchical supervision framework to explicitly model and trade off PoA and its associated cost.

The principle of hierarchical supervision in organizations has been widely adopted in previous studies. For example, Zhang \emph{et al}. developed an organization-based control framework to speed up multiagent learning in a network of agents \cite{zhang2009integrating,zhang2010self}. Bazzan et al. resorted to the hierarchical supervision as a social instrument for solving the prisoner's dilemma \cite{bazzan2011learning}. In this work, we propose the methodology to realize general hierarchical supervision in organizations and focus on seeking the optimal control solution to trading off between cost and performance in such organizations.

There is also tremendous amount of work that aims to solve coordination issues in the two case problems. For the NL problem, numerous mechanisms have been proposed for efficient emergence of social norms while agents interact with each other using learning (particularly reinforcement learning) methods. These mechanisms include the social learning strategy~\cite{sen2007emergence,airiau2014emergence,chen2018social}, the collective interaction protocol~\cite{yu2014collective,hao2017efficient}, the utilization of topological knowledge~\cite{villatoro2009topology,tang2018optimal,hasan2018context} and agents' observation capabilities~\cite{villatoro2011social,vouros2017learning}. Various kinds of solutions have also been proposed to solve the \emph{selfish equilibria} problem in MARS, or more broadly, social dilemmas, when agents use rational learning strategies for interaction. For example, Bazzan \emph{et al.} resorted to social instruments of hierarchy and coalition to promote cooperation in Iterated Prisoner's Dilemmas \cite{bazzan2011learning}. Oh and Smith applied social learning to promote the social welfare in MARS problems \cite{oh2008few}. Our work supplements the literature by providing new effective solutions to these two challenging problems. This is also claimed as the main contribution of this paper.

\section{Conclusions}\label{sec:conclusions}

In this paper, we argue for the benefits of considering both centralized control and decentralized interactions in solving a coordination problem. By trading off between these two aspects, an optimal middle ground solution, represented by the lowest PoG, can be discovered. Two explicit case studies are presented to exemplify the implementation of the hierarchical supervision framework: the NL problem is a pure coordination game where agents' behavior should keep consistently such that a global social norm can be established, while the MARS problem is a typical social dilemma problem where agents should reason in a bounded rational manner to avoid tragedy of the commons. Simulation results show that the proposed framework is capable of promoting coordination among agents while bounding the administrative cost to a minimum value. As such, the hierarchical supervision framework and its implementation methodology are indeed effective mechanisms towards coordination in organizational control.

The work in this paper inspires several directions for future work. First, in this paper we implement the framework using simulations, that is, sampling methods are used to fit the PoA and PoM relationship curve before computing the lowest PoG. This proof-of-concept validation is reasonable for the two case problems, particularly for the NL problem where theoretical analysis and proof over the solution and its performance are still open issues in this area. However, building on the rich work in theoretical analysis on PoA \cite{wang2016analysis,roughgarden2015intrinsic}, it is possible to derive the optimal PoG in a closed form by introducing a cost function into the problem formulation and solving the optimization problem given by Eq.~(\ref{equ:optimization}) directly. Second, richer phenomenon or more important discoveries may be revealed under various problem settings or specifications, e.g., combination functions, or cost measurement etc. Last, we are expecting more implementations of this framework and its methodology in solving other real-world coordination problems, in which efficiency and cost are the two main optimized objectives (e.g., resources allocation in cognitive radio networks). These issues are left for the future work.


\bibliographystyle{ACM-Reference-Format}  
\bibliography{PoG-aamas19}  

\end{document}